\documentstyle[psfig,amstex]{mn}
\def\la{Ly$\alpha$}
\def\etal{et~al.}
\def\spose#1{\hbox to 0pt{#1\hss}}
\def\lta{\mathrel{\spose{\lower 3pt\hbox{$\mathchar"218$}}
     \raise 2.0pt\hbox{$\mathchar"13C$}}}
\def\gta{\mathrel{\spose{\lower 3pt\hbox{$\mathchar"218$}}
     \raise 2.0pt\hbox{$\mathchar"13E$}}}

\def\clean{{\sc clean}}

\def\uv{{\it uv}}

\def\aips{{\sc aips}}
\def\robust{{\sc robust}}
\def\imagr{{\sc imagr}}

\title[]{Deep radio observations of 3C324 and 3C368: evidence for
jet--cloud interactions} 

\author[P.~N.~Best \etal]{P.~N.~Best$^1$, C.~L.~Carilli$^2$,
S.~T.~Garrington$^3$, M.~S.~Longair$^4$ and\\
\vspace*{0.5mm}
{\LARGE H.~J.~A.~R\"ottgering$^1$}\\  
$^1$ Sterrewacht Leiden, Huygens Laboratory, P.O.Box 9513, 2300 RA 
Leiden, The Netherlands\\ 
$^2$ NRAO, P.O.Box 0, Socorro, NM 87801-0387, USA\\
$^3$ The University of Manchester, NRAL, Jodrell Bank, Lower Withington,
Macclesfield, Cheshire, SK11 9DL, England\\ 
$^4$ Cavendish Labs, Madingley Road, Cambridge, CB3 0HE, England}
\pagerange{\pageref{firstpage}--\pageref{lastpage}}
\pubyear{1996}
\begin{document}
\label{firstpage}

\maketitle

\begin{abstract}
\noindent High resolution, deep radio images are presented for two distant
radio galaxies, 3C324 ($z=1.206$) and 3C368 ($z=1.132$), which are both
prime examples of the radio--optical alignment effect seen in powerful
radio galaxies with redshifts $z \gta 0.6$. Radio observations were made
using the Very Large Array in A--array configuration at 5 and 8\,GHz, and
using the MERLIN array at 1.4 and 1.65\,GHz. Radio spectral index, radio
polarisation, and rotation measure maps are presented for both sources.

\noindent Radio core candidates are detected in each source, and by
aligning these with the centroid of the infrared emission the radio and
the optical\,/\,infrared images can be related astrometrically with 0.1
arcsec accuracy. In each source the radio core is located at a minimum of
the optical emission, probably associated with a central dust lane. Both
sources also exhibit radio jets which lie along the directions of the
bright strings of optical knots seen in high resolution Hubble Space
Telescope images. The northern arm of 3C368 shows a close correlation
between the radio and optical emission, whilst along the jet direction of
3C324 the bright radio and optical knots are co--linear but not
co--spatial. These indicate that interactions between the radio jet and
its environment play a key role in producing the excess ultraviolet
emission of these sources, but that the detailed mechanisms vary from
source to source.

\noindent 3C368 is strongly depolarised and has an average rest--frame
rotation measure of a few hundred rad\,m$^{-2}$, reaching about
1000\,rad\,m$^{-2}$ close to the most depolarised regions.  3C324 has
weaker depolarisation, and an average rest--frame rotation measure of
between 100 and 200\,rad\,m$^{-2}$. Both sources show large gradients in
their rotation measure structures, with variations of up to
1000\,rad\,m$^{-2}$ over distances of about 10\,kpc.

\end{abstract}


\section{Introduction}
\label{intro}

The discovery that the optical, ultraviolet and line emission of powerful
high redshift radio galaxies is elongated and aligned along the direction
of the radio axis (McCarthy \etal\ 1987, Chambers \etal\
1987)\nocite{mcc87,cha87}, a phenomenon known as the `alignment effect',
indicates a close association between radio source activity and the
optical emission of these galaxies. A number of processes have been
proposed to account for these radio--optical correlations (see McCarthy
1993 for a review),\nocite{mcc93} the most promising models falling into
two broad categories: interaction models in which the correlations arise
directly through interactions between the radio jet and its environment,
and illumination models in which radiation from a partially obscured
active galactic nucleus (AGN) both photo--ionises the emission line gas
within an ionisation cone and is scattered by dust grains or electrons
producing the optical and ultraviolet alignment.

One of the earliest models for the alignment effect, which has remained
popular, is star--formation induced by the passage of the radio jets
(e.g. Rees 1989).\nocite{ree89} This process has been observed in
jet--cloud interactions at low redshifts, for example in Minkowski's
object \cite{bro85,bre85} and in the radio lobe of 3C285
\cite{bre93}. There is quite convincing evidence that it also occurs in at
least some distant radio galaxies (e.g. Best \etal\ 1997a, Dey \etal\
1997)\nocite{bes97b,dey97} but, as yet, direct evidence for the presence
of young stars in the aligned structures is limited.

A large proportion of the powerful distant radio galaxies which have been
studied using spectropolarimetry have optical emission which is polarised
at up to the 15\% level, and scattered broad lines in their spectra
(e.g. Dey \etal\ 1996, Cimatti \etal\ 1996, 1997, and references
therein)\nocite{cim96,dey96,cim97a}.  A significant proportion of the
aligned emission of these galaxies must therefore be scattered light from
an obscured AGN. In an optically unbiased sample of distant radio
galaxies, however, Tadhunter \etal\ \shortcite{tad97} detected significant
polarisation in only 40\% of the galaxies although most showed large
ultraviolet excesses.

Nebular continuum emission provides another important alignment
mechanism. Dickson \etal\ \shortcite{dic95} showed that a combination of
free--free, free--bound, and two--photon continuum, as well as the Balmer
forest, can contribute a significant proportion (5--40\%) of the
ultraviolet emission from the nuclear regions of powerful radio galaxies,
and this may be even higher in the extended aligned emission regions
(e.g. Stockton \etal\ 1996)\nocite{sto96a}.  Since either or both of 
photo--ionisation by the AGN and shock--ionisation by the radio jet may be
responsible for exciting the gas, distinguishing between the jet--cloud
interaction model and the illumination model is made more difficult.

The structure and properties of the radio emission from these distant
radio galaxies can offer important clues to the nature of the
radio--optical correlations.  Hubble Space Telescope (HST) observations of
28 radio galaxies at redshifts $z \sim 1$, selected from the revised 3CR
catalogue of Laing \etal\ \shortcite{lai83}, have shown that for the
majority of small radio sources the alignment effect manifests itself in
the form of strings of bright knots tightly aligned between the two radio
lobes (Best \etal\ 1996, 1997b).\nocite{bes96a,bes97c} Deep, high
resolution radio observations of these sources provide a critical test of
the interaction\,/\,illumination models: if the locations of the radio
jets and the strings of optical activity are co--linear, then the
influence of the radio jets must play a key role in the alignment
effect. This would not, however, provide direct proof of the jet--induced
star formation hypothesis: the emission may be nebular continuum emission
from warm gas shocked by the radio jets, or may be associated with an
increased scattering efficiency in these regions, for example by jet
shocks breaking up cold gas clouds and exposing previously hidden dust
grains, thereby increasing the surface area for scattering along the jet
axis \cite{bre96b}.

One of the major problems in interpreting the radio--optical structures of
distant radio galaxies has been accurate relative astrometry of the radio
and optical\,/\,infrared frames of reference. Radio observations are
automatically in the International Celestial Reference Frame (ICRF), with
positional errors of about 0.01 arcsec, but the absolute positions of the
HST images are uncertain at the 1 arcsecond level \cite{lat97}. Such
astrometric uncertainties mean that bright radio knots may be either
correlated or anti--correlated with the position of the luminous optical
emission regions. This problem can be alleviated somewhat by
multi--frequency radio observations of sufficient sensitivity to detect a
flat--spectrum central radio core: this core is expected to be coincident
with the nucleus of the galaxy, which itself is located roughly at the
centroid of the infrared image of the galaxy. The optical (rest--frame
ultraviolet) emission cannot be used for this procedure, since it is often
severely affected by both dust extinction (e.g. de Koff \etal\ 1996, Best
\etal\ 1997b)\nocite{bes97c,dek96} and the aligned emission. The infrared
images also contain a component of aligned emission (e.g. Eisenhardt \&
Chokshi 1990; Rigler \etal\ 1992; Dunlop \& Peacock 1993; Best \etal\
1997b,1998)\nocite{eis90,rig92,dun93,bes98d} but at a significantly lower
level, and are seen to show a sharp central peak which can be
astrometrically aligned with the radio core. 

It should be noted that this process can never be 100\% reliable, owing to
the ambiguity of identifying radio core components based upon spectral
index information alone; the case of 3C356 (Fernini \etal\ 1993, Best
\etal\ 1997a) is a case in point. For the radio galaxies in the current
paper, however, further radio--optical correlations are observed when this
process is employed (see Sections 3 and 4), indicating that the core
identifications are most likely correct.

The properties of the radio emission also provide a probe of the
large--scale physical environment of the radio source. There has been
growing evidence that powerful radio galaxies at large redshifts lie in
rich (proto--) cluster environments, based upon measures of galaxy
cross--correlation functions (e.g. Yates \etal\ 1989)\nocite{yat89},
detections of powerful extended X--ray emission from the vicinity of
distant radio sources (e.g. Crawford and Fabian 1996)\nocite{cra96b},
detections of companion galaxies in narrow--band images and with
spectroscopy (e.g. McCarthy 1988, Dickinson \etal\
1997)\nocite{mcc88,dic97a}, and the very bright infrared magnitudes and
large characteristic sizes of the radio galaxies themselves, suggesting
that they are as large and luminous as brightest cluster galaxies (see
Best \etal\ 1998 for a review)\nocite{bes98d}. Radio data provide further
evidence in support of this hypothesis. At low redshifts, radio galaxies
in rich cluster environments, such as Cygnus A, tend to have very large
rotation measures, $RM \gta 1000$\,rad\,m$^{-2}$ \cite{dre87,tay94}.
Carilli \etal\ \shortcite{car97} have studied a sample of 37 radio
galaxies with redshifts $z > 2$, and detect rotation measures exceeding
1000\,rad\,m$^{-2}$ in 19\% of them. They consider this percentage to be a
conservative lower limit, and interpret these large rotation measures as
arising from hot, magnetised (proto--)cluster atmospheres surrounding the
sources, with field strengths $\sim 1$\,nT.  Detailed studies of
individual sources with very high rotation measures support this
interpretation \cite{car94a,pen97}. In comparison with these very distant
sources, the polarisation properties of radio galaxies with redshift $z
\sim 1$ have not been well studied at high angular resolution (although
see Pedelty \etal\ 1989, Liu and Pooley 1991, Fernini \etal\ 1993, and
Johnson \etal\ 1995)\nocite{ped89a,joh95,fer93,liu91a}.

We have selected two radio galaxies from the 3CR sample, with redshifts in
excess of one: 3C324 ($z=1.206$) and 3C368 ($z=1.132$). These galaxies are
both prime examples of the alignment effect; their Hubble Space Telescope
(HST) images show highly extended optical morphologies with strings of
bright knots close to the radio axis \cite{lon95}. We have carried out
sensitive, multi--frequency, polarimetric radio observations of these
radio sources at high angular resolution using the VLA and MERLIN. A
description of the observations and the data reduction techniques is given
in Section~\ref{observs}. In Section~\ref{sec3c324}, we present the radio
data for 3C324, and make a comparison with the optical and infrared
images. This is repeated for 3C368 in Section~\ref{sec3c368}. These
sources are compared in Section~\ref{concs}, in which we also summarise
our conclusions.  Throughout the paper we assume $H_0 =
50$\,km\,s$^{-1}$\,Mpc$^{-1}$ and $\Omega = 1$, and all positions are
given in equinox J2000 coordinates.

\section{Observations and data reduction}
\label{observs}

\subsection {Very Large Array (VLA) observations}
\label{vla}

Observations of 3C324 and 3C368 were made in July 1995, in the 5\,GHz and
8\,GHz bands of the VLA in A--array configuration. 3C368 was also observed
in the 1.4\,GHz band. Details of the observations are given in
Table~\ref{vlatab}.

\begin{table*}
\begin{center}
\begin{tabular}{lcccccrrr}
Source&Redshift&Observation & Interferometer & Frequencies & Bandwidth & Integration & Map\hspace{4.0mm} & Restoring beam \\
     &     &  Date      &                &      &  &Time\hspace{4.0mm} &  rms noise\hspace{1.0mm} & FWHM\hspace{5.8mm} \\
     &     &              &      &[MHz]&[MHz]&[min]\hspace{4.0mm} &[$\mu$Jy/beam]&[arcsec]\hspace{6.0mm} \\
3C324&1.206&11th July 1995& VLA & 4535, 4885      & 50 & 77\hspace{5.0mm}  & 26\hspace{5.0mm} &0.40\hspace{8.0mm} \\
     &     &              & VLA & 8085, 8335      & 50 & 291\hspace{5.0mm} & 13\hspace{5.0mm} &0.22\hspace{8.0mm} \\
3C368&1.132&31st July 1995& VLA & 1465, 1515      & 50 & 120\hspace{5.0mm} & 51\hspace{5.0mm} &1.35\hspace{8.0mm} \\
     &     &              & VLA & 4535, 4885      & 50 & 60\hspace{5.0mm}  & 25\hspace{5.0mm} &0.45\hspace{8.0mm} \\
     &     &              & VLA & 8085, 8335      & 50 & 123\hspace{5.0mm} & 20\hspace{5.0mm} &0.23\hspace{8.0mm} \\
     &     &3rd March 1996& MERLIN & 1658 & 15$\times$1& 780\hspace{5.0mm} &119\hspace{5.0mm} &0.23$\times$0.15\hspace{4.0mm} \\ 
     &    &16th March 1996& MERLIN & 1420 & 15$\times$1& 780\hspace{5.0mm} &150\hspace{5.0mm} &0.25$\times$0.16\hspace{4.0mm} \\ 
\end{tabular}
\end{center}
\caption{\label{vlatab} The parameters of the radio observations.}
\end{table*}

The observations were made using standard VLA procedures. Primary flux
calibration was achieved using the bright source 3C286, and the nearby
secondary calibrators 1513+236 and 1824+107 were observed regularly to
provide accurate phase calibration for 3C324 and 3C368 respectively. The
observations of these calibrators, spaced over a wide range of parallactic
angles, also determined the on--axis antenna polarisation response
terms. Scans of 3C286, separated in time by 6 hours, provided absolute
polarisation position angle calibration. The uncertainty in the
calibration of the position angles, estimated from the difference between
the solutions for the scans of 3C286, was about $\pm$2 degrees at each
frequency. For observations at 4710 and 8210\,MHz, this corresponds to an
uncertainty in the absolute value of the rotation measure of about
20\,rad\,m$^{-2}$.
        
The data were reduced using the \aips\ software provided by the National
Radio Astronomy Observatory. The two IFs of the 8\,GHz band data were
combined to provide a single dataset at 8210\,MHz and those in the
1.4\,GHz band were combined at 1490\,MHz. The different IFs of the 5\,GHz
band data were reduced separately. The data were \clean ed, and phase
self--calibration was used to improve the map quality. The calibration was
generally very successful, yielding images with off--source noise levels
within 50\% of the expected theoretical values.

\subsection{MERLIN observations}
\label{merlin}

3C368 was observed by MERLIN at 1420 and 1658\,MHz, with the parameters of
the observations provided in Table~\ref{vlatab}. The run at 1420\,MHz was
made without the Wardle telescope, and the Mk2 telescope was used as the
home--station telescope at Jodrell Bank. The total track length of each
run was 13-hours, and the data were recorded in 15 $\times$ 1\,MHz
bandwidth channels.

Initial amplitude and bandpass calibration were carried out using 2134+004
and 0552+398, whose flux densities were determined by comparison with
3C286. Further phase and amplitude calibration was done using 1752+119,
which was observed for 2 minutes every 10 minutes throughout the run and
was assumed to be unresolved. This source was also used to determine the
instrumental polarisation to within 0.5\%. A short observation of 3C286
provided absolute polarisation position angle calibration.

Images of 3C368 were then made separately at each of the two frequencies,
and several cycles of self--calibration in AIPS and also in DIFMAP
\cite{she97} were used to improve the final image quality.

\subsection{Images}
\label{images}

For both sources, images were made at full angular resolution in the
Stokes parameters I, Q and U at each frequency, by \clean ing the final
datasets using the \aips\ task \imagr\ with the data weighting parameter
\robust\ set to zero.  The FWHM of the Gaussian restoring beams are listed
in Table~\ref{vlatab}.  The Stokes I images of the two IFs in the 5\,GHz
band were combined to produce a single 4710\,MHz total intensity image.

To produce maps of spectral index, rotation measure, and depolarisation,
observations at different frequencies must be matched in their
\uv-coverage and beam size. Images of the 8\,GHz data were therefore
created at the lower resolution of the 5\,GHz data, by applying an upper
cut-off in the \uv\ data matching the longest baseline sampled at 5\,GHz,
together with \uv\ tapering to retain the smooth coverage of the \uv\
plane. The \clean ed components were restored using a Gaussian beam
appropriate for the 5\,GHz data. The resulting images were then checked
against the 5\,GHz images for any astrometric differences between the
positions of the hotspots, resulting from the self--calibration
procedure. No offsets in excess of 0.1 arcsec were found.

\begin{table*}
\begin{tabular}{lcccccccccc}
Source&Region&  Total  &Fractional  &  Total  &Fractional  &  Total  &Fractional  &Spec.   & Depolar. &Rotation \\
      &      &intensity&Polarisation&intensity&Polarisation&intensity&Polarisation& index  & measure  &measure  \\
      &      &8210\,MHz& 8210\,MHz  &4710\,MHz& 4710\,MHz  &1490\,MHz& 1490\,MHz  &$\alpha$&$DM_{8.2}^{4.7}$&$RM$\\
      &      & [mJy]   & [\%]       & [mJy]   & [\%]       & [mJy]   & [\%]       &        &     &[rad\,m$^{-2}$]\\
3C324 &Whole &  332    & 18.1       & 686     &  15.8      &  ---    &  ---       & 1.31   &  0.88    &   41    \\
      &W lobe&  93     & 17.2       & 196     &  15.1      &  ---    &  ---       & 1.34   &  0.87    &   45    \\
      &E lobe&  238    & 17.8       & 489     &  15.6      &  ---    &  ---       & 1.30   &  0.88    &   38    \\
      &Core  &  0.14   & ---        & 0.17    &  ---       &  ---    &  ---       & 0.30   &  ---     &  ---    \\
      &W knot&  0.74   & ---        & 1.00    &  ---       &  ---    &  ---       & 0.54   &  ---     &  ---    \\
3C368 &Whole &  98     & 9.6        & 229     &  6.7       &  1205   &  1.1       & 1.52   &  0.70    &   85    \\
      &N lobe&  47     & 9.9        & 109     &  8.2       &   564   &  0.6       & 1.51   &  0.83    &  100    \\
      &S lobe&  50     & 9.8        & 118     &  5.4       &   641   &  1.5       & 1.55   &  0.55    &   63    \\
      &Core  &  0.16   & ---        & 0.22    &  ---       &   ---   &  ---       & 0.54   &  ---     &  ---    \\
      &S knot&  0.28   & ---        & 0.55    &  ---       &   ---   &  ---       & 1.22   &  ---     &  ---    \\
\end{tabular} 
\caption{\label{radprops} Global properties of the various components of
the two sources. The spectral index and depolarisation measures listed are
those calculated between 4710 and 8210\,MHz. The rotation measures are the
mean values for a region of the source as calculated from the three
frequencies 4535, 4885 and 8210\,MHz. The `global' depolarisation measures
are calculated by first determining the scalar fractional polarisation at
the two frequencies and then dividing.}
\end{table*}

Spectral index maps (where we define the spectral index, $\alpha$, as
$I_{\nu} \propto \nu^{-\alpha}$) were made for both sources using these
8210\,MHz matched--resolution total intensity maps and the combined
4710\,MHz total intensity maps.  Spectral index values were only
calculated for regions of the images with surface brightnesses in excess of
5$\sigma$ at both frequencies, where $\sigma$ is the rms noise level of
the image. Rotation measures were derived using the position angle of the
polarised intensity at the three frequencies 4535, 4885 and 8210\,MHz.
Rotation measures were only derived in regions of the source for which
polarised intensity was detected in excess of 4$\sigma$ at all three
frequencies. The fractional polarisation at each frequency was derived by
dividing the flux density determined from the total polarised intensity
map, after correction for noise bias, by the flux density of the total
intensity map, and is therefore a scalar rather than vector average of the
polarisation. The depolarisation measure, $DM^{4.7}_{8.2}$, defined as the
ratio of the fractional polarisation at 4710\,MHz to that at 8210\,MHz,
was derived only in regions of the source where the polarised intensity in
both images exceeded $4\sigma$.

\section{3C324}
\label{sec3c324}

\subsection{Radio properties of the source}
\label{324radio}

\begin{figure*}
\centerline{
\psfig{figure=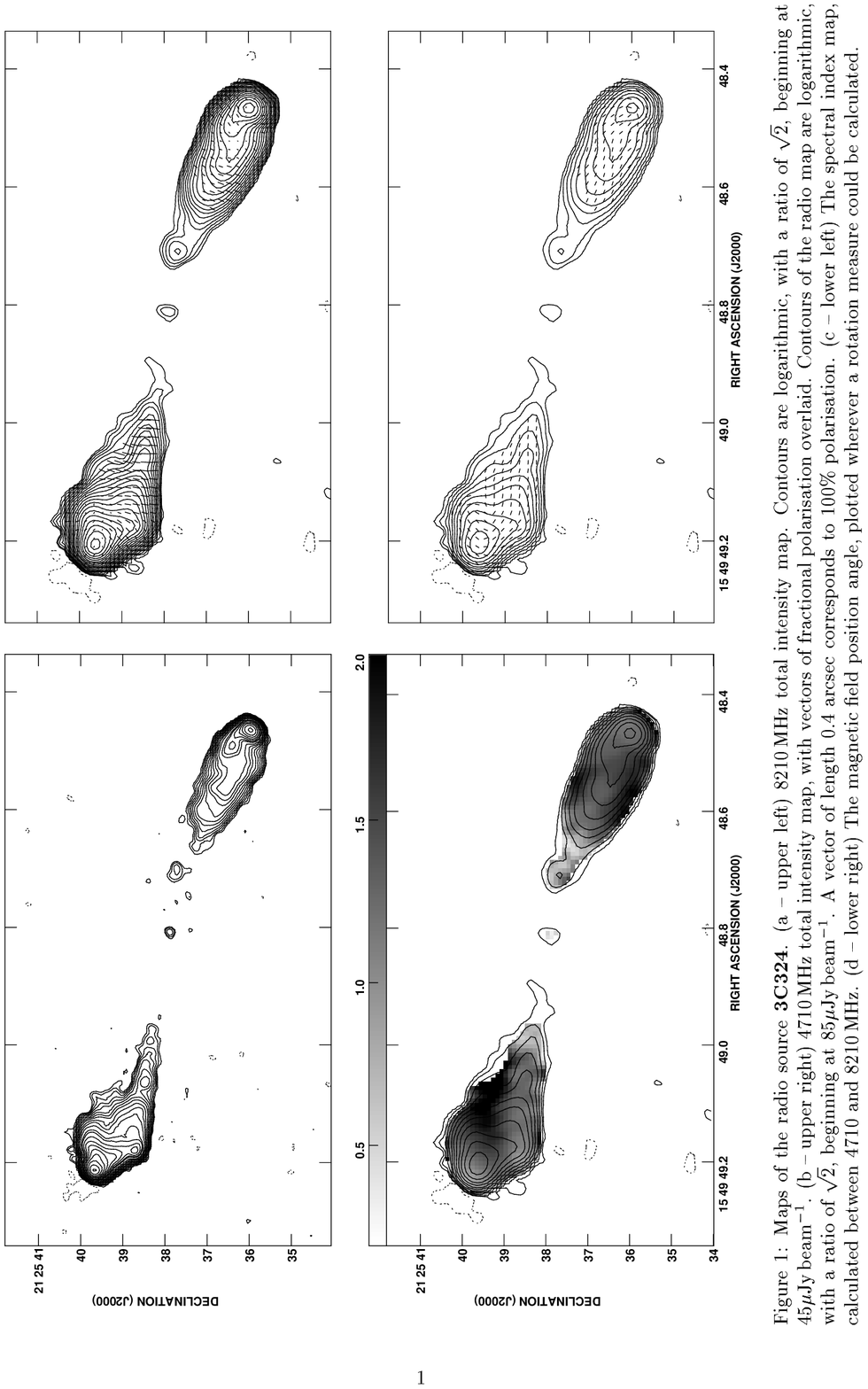,clip=,height=\textheight}
}
\end{figure*}

\begin{figure*}
\centerline{
\psfig{figure=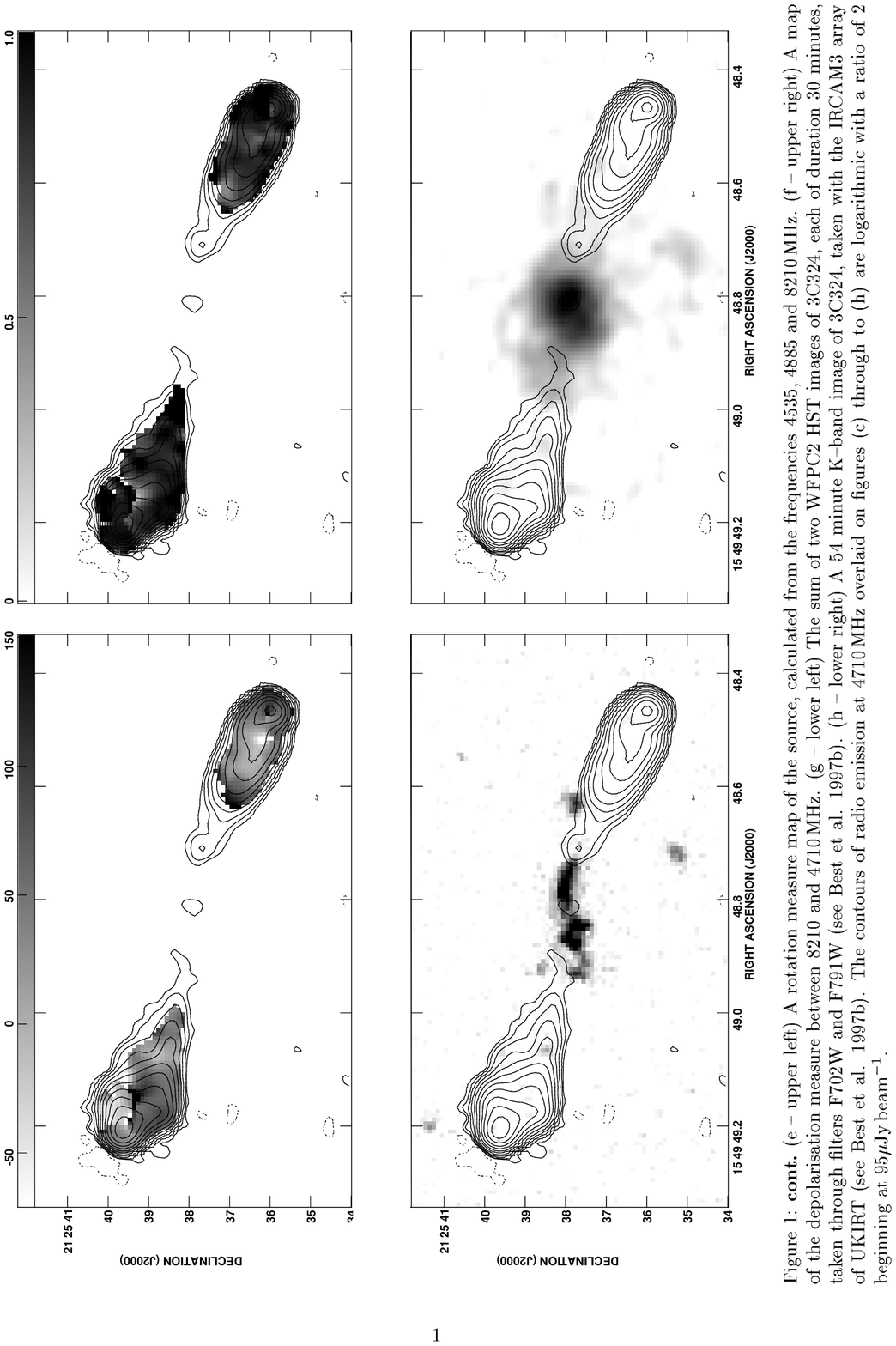,clip=,height=\textheight}
}
\end{figure*}

\addtocounter{figure}{2}

Figure~1a shows the 8210\,MHz total intensity contour map of 3C324, and
Figure~1b shows the corresponding 4710\,MHz total intensity map with
vectors of the radio polarisation overlaid. A grey--scale map of the
spectral index calculated between these two frequencies, as described in
Section~\ref{images}, is shown in Figure~1c with contours of the 4710\,MHz
radio emission overlaid. The source can be separated into four components:
the western lobe, the eastern lobe, a central knot, and a knot close to
the western lobe. The integrated radio properties of these four regions
are listed separately in Table~\ref{radprops}. The central knot has a
spectral index of 0.30, significantly flatter than the rest of the
source. Although the cores of low redshift radio sources tend to have very
flat spectra, those seen in radio sources with redshifts $z \gta 1.5$
often have somewhat steeper spectra \cite{lon93,car97,ath97}, and so a
spectral index of 0.3 for a radio core at redshift $z \sim 1$ is not
unreasonable.  We therefore identify this knot with the active galactic
nucleus.

In the high angular resolution observations at 8\,GHz, a radio jet is
detected passing through a series of knots along the southern edge of the
eastern lobe and then making a dramatic 55$^{\circ}$ turn to the north at
the eastern end of the lobe. In Figure~1d, we show a map of the magnetic
field position angle; these vectors lie parallel to the jet direction
along this string of knots, and echo the sharp turn northwards (see also
the polarisation position angle in Figure~1b). The relatively flat
spectral index of this region of the radio source (Figure~1c) adds further
support to this picture. The north--western region of this eastern lobe
has a steep spectral index, characteristic of ageing lobe electrons
\cite{sch68}.

To the west of the core no clear radio jet is detected, although the knot
of radio emission along this axis has a relatively flat spectral index,
$\alpha \approx 0.54$, and may be a jet knot. This knot lies directly upon
a line from the jet in the eastern lobe through the radio
core. Interestingly, the hotspots of the western and eastern lobes and the
radio core are approximately co-linear, although this line is misaligned
by 10$^{\circ}$ from the current axis of the radio jet.

In Figure~1e, a map of the rotation measure is shown, calculated as
described in Section~\ref{images}. The mean rotation measures of the
eastern and western lobes are 38\,rad\,m$^{-2}$ and 45\,rad\,m$^{-2}$
respectively, only about twice the 1$\sigma$ limit for a secure detection
(see Section~\ref{vla}).  There is, however, considerable structure in the
rotation measure map, the rotation measure changing from about $-50$ to
about +150\,rad\,m$^{-2}$ over an angular scale of order one arcsecond;
such gradients are unlikely to be of Galactic origin \cite{lea87}. Thus,
the observed values for the rotation measures must be scaled up by a
factor of $(1+z)^2$, or 4.87, and so the variations in the rest--frame
rotation measure for 3C324 approach 1000\,rad\,m$^{-2}$ over distances of
about 10\,kpc.

The regions of the highest rotation measure are associated with regions in
which the strongest depolarisation is observed between 8210\,MHz and
4710\,MHz in Figure~1f. The observed values for the depolarisation and
rotation measures can be compared if it is assumed that the depolarisation
results from a Gaussian distribution of Faraday depths within each
beamwidth. If the standard deviation of that Gaussian distribution is
$\Delta$, measured in rad\,m$^{-2}$, then the polarisation, $m_{\lambda}$
at a given wavelength is given by the depolarisation law of Burn
\shortcite{bur66}:

\begin{equation}
m_{\lambda} = m_{\rm 0} {\rm exp}[-2 \Delta^2\lambda^4].
\label{depollaw}
\end{equation}

\noindent For observations at 4710 and 8210\,MHz, this gives

\begin{equation}
\Delta = 185 (1+z)^2 \left[ -{\rm ln} DM_{8.2}^{4.7} \right]^{0.5}
\label{deltaeqn}
\end{equation}

\noindent which for 3C324 yields $\Delta \sim 320$\,rad\,m$^{-2}$,
comparable to the range of rotation measures observed. These correlations
between the depolarisation and the gradient of the rotation measures
suggest that the Faraday screen is external, arising from gas surrounding
the radio lobes.

\subsection{Radio--optical correlations}
\label{324radopt}

In Figure~1g, we show an HST image of 3C324, with the radio contours from
the 4710\,MHz observation overlaid. The HST image represents the sum of
two 30 minute observations, taken through the F702W and F791W filters of
WFPC2 \cite{lon95}. In Figure~1h, we show a 54-minute K--band image of the
source taken using IRCAM3 of UKIRT in 1 arcsecond seeing. The HST and
UKIRT images were astrometrically aligned using a number of unresolved
objects appearing on both frames. These were then aligned with the radio
data assuming that the radio core is coincident with the centre of the
infrared image, providing an astrometric accuracy of approximately 0.1
arcseconds. The details of these HST and UKIRT images have been
extensively discussed elsewhere \cite{bes97c}; here we consider only their
relation to the new radio data.

The most striking feature of Figure~1g is that the regions of bright radio
and optical emission are co-linear, but not co-spatial. The radio core is
located at a minima in the optical emission, probably associated with a
central dust lane of extinction $E(B-V)\gta0.3$ \cite{lon95,dic96}. To the
west of this, a string of bright optical knots curve round towards the
western radio knot; at the end of this curve they point directly towards
the western radio hotspot. There is no bright optical emission coincident
with the radio knot, but beyond this there lies a further optical knot.
In the eastern arm of the radio source, the optical knots do not form such
a smooth structure, but the radio jet becomes luminous where the bright
optical emission ceases. The apparent rotation of the axis of 3C324 with
decreasing size, from that defined by the radio hotspots, through that
defined by the radio jet, to that defined by the optical emission has been
interpreted by Dickinson \etal\ \shortcite{dic96} as being associated with
precession of the axis of the AGN. The sharp bend of the radio jet in the
eastern arm would then represent a location where the jet makes a glancing
impact upon the cocoon wall \cite{cox91}.

The UKIRT image (see also the K--band map of Dunlop and Peacock
1993)\nocite{dun93} has significantly lower angular resolution but,
intriguingly, it shows faint emission in the eastern radio lobe close to
the region which shows a large gradient in the rotation measure and where
the depolarisation is strongest. These features may be associated with a
companion galaxy to 3C324. A deeper, high resolution image is required to
confirm this.

\section{3C368}
\label{sec3c368}
\subsection{Radio properties of the source}
\label{368radio}

\begin{figure*}
\centerline{
\psfig{figure=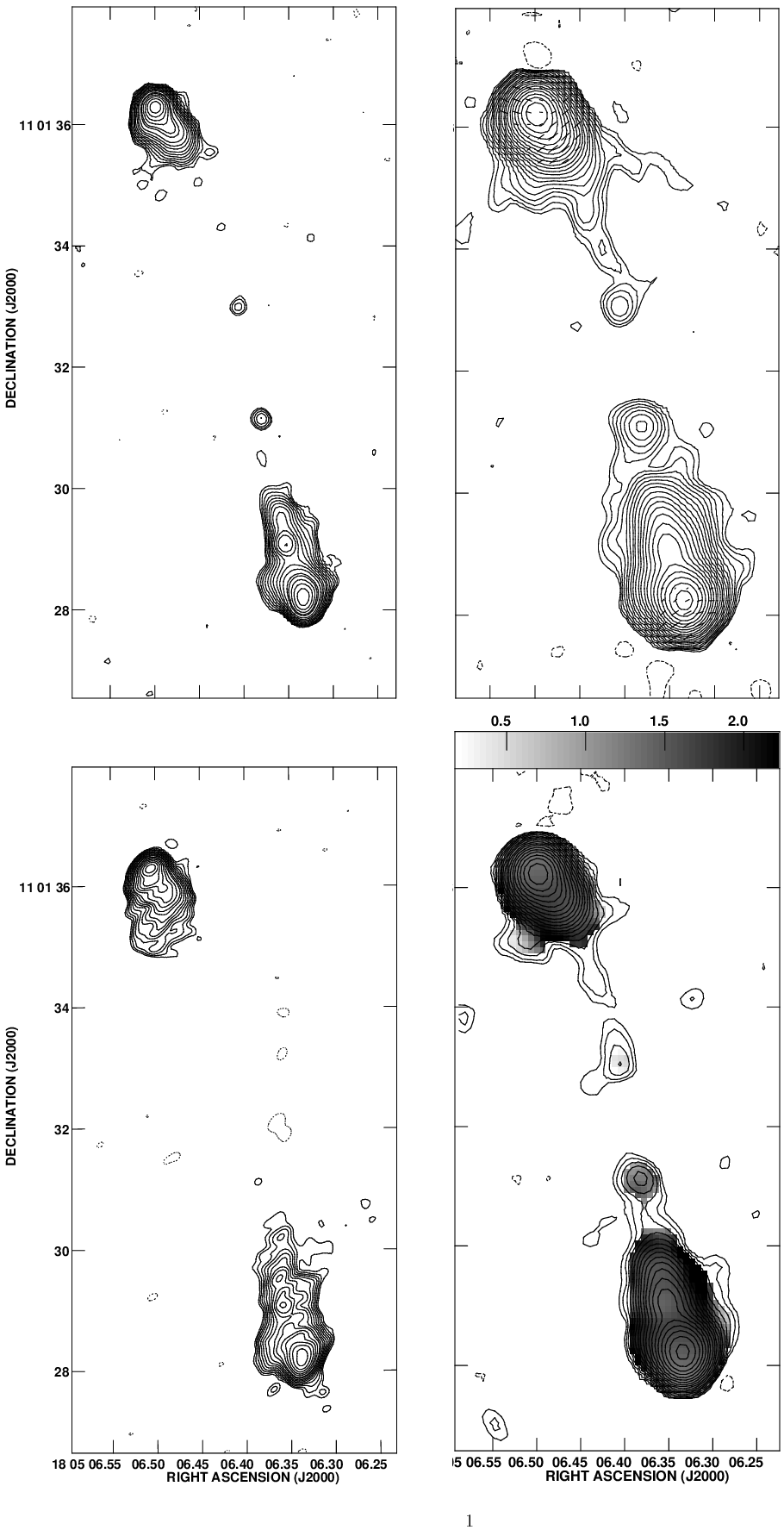,clip=,height=\textheight}
}
\end{figure*}

\begin{figure*}
\centerline{
\psfig{figure=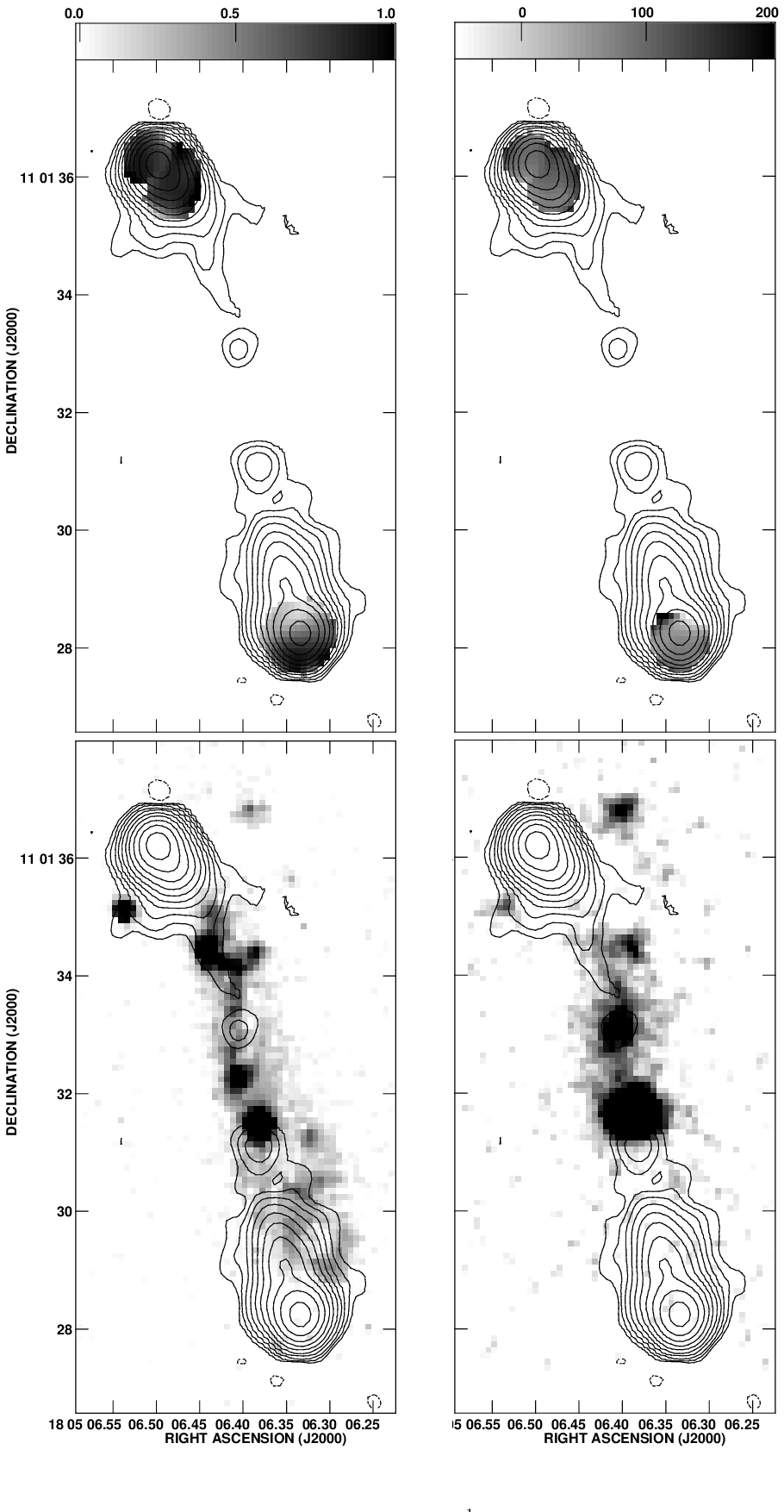,clip=,height=\textheight}
}
\end{figure*}

\begin{figure*}
\centerline{
\psfig{figure=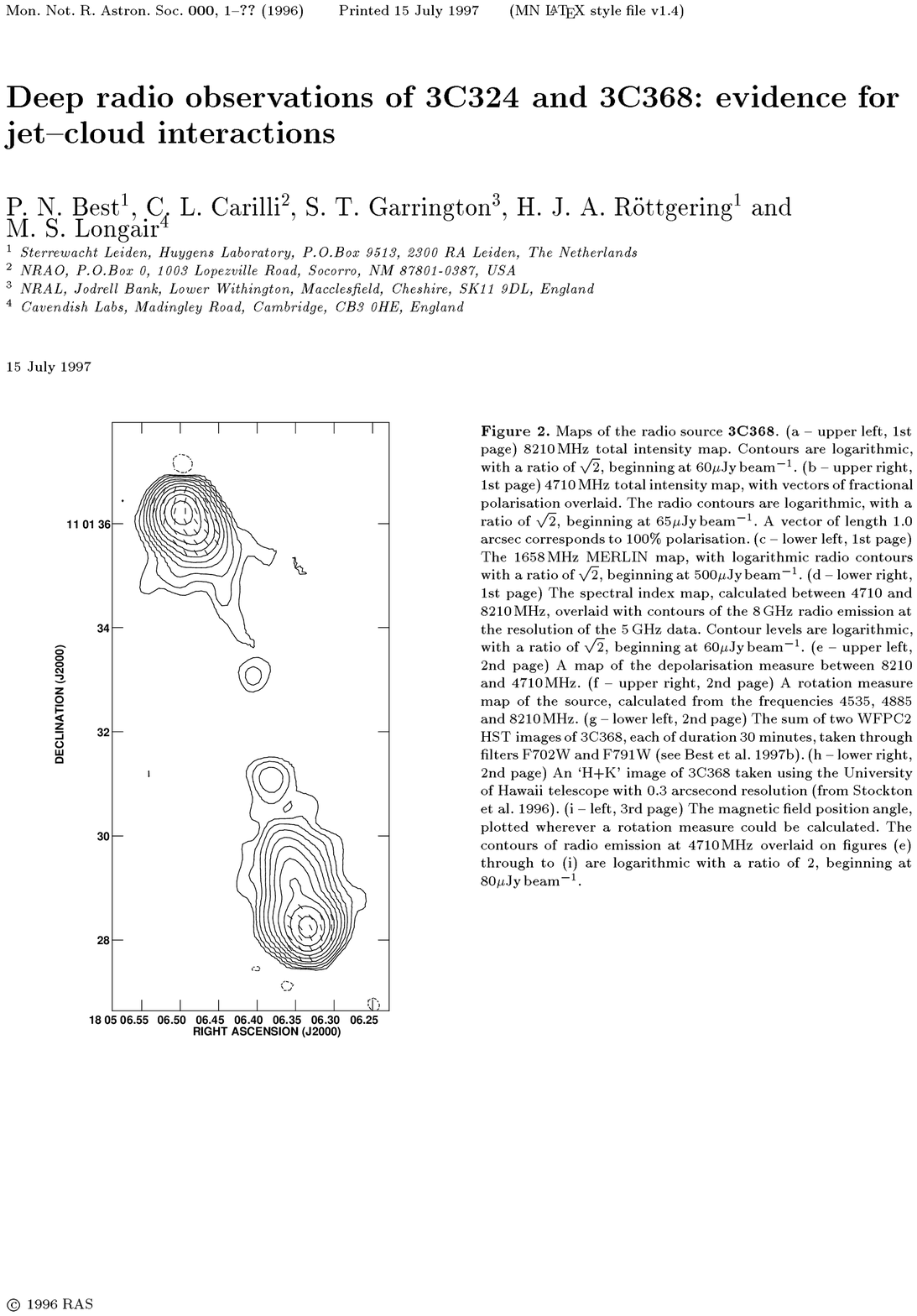,clip=,width=\textwidth}
}
\end{figure*}
\nocite{bes97c,sto96a}

Figure~2a shows the VLA 8210\,MHz total intensity contour map
of 3C368, and Figure~2b the corresponding 4710\,MHz intensity
map. A MERLIN 1658\,MHz map is given in Figure~2c.
Figure~2d shows a grey--scale map of the spectral index between
4710 and 8210\,MHz, overlaid with a contour map of the 8\,GHz radio
emission made at the lower resolution of the 5\,GHz data (see
Section~\ref{images}).

Figure~2a shows two unresolved central components. The integrated
properties of each of these knots and of the northern and southern radio
lobes are listed in Table~\ref{radprops}. The southern of these two knots
has been previously detected by Djorgovski \etal\ \shortcite{djo87} and
Chambers \etal\ \shortcite{cha88}, and associated by these authors with
the radio core. However, our observations indicate that the spectral index
of this knot is $\alpha \approx 1.22$, much steeper than would be expected
of a radio core. The northern knot, detected here for the first time, has
a much flatter spectral index, $\alpha \approx 0.54$, more consistent with
the core spectral indices that might be expected at this redshift (see the
discussion in Section~\ref{324radio}). Furthermore, the 5\,GHz and lower
resolution 8\,GHz radio maps (Figures~2b and 2d) show a clear jet
structure leading from this northern knot to the northern radio lobe,
supporting this hypothesis. In the following discussion, we assume the
northern knot is associated with the AGN.

Overlaid upon Figure~2b, we show the polarisation vectors of the source at
5\,GHz. Perhaps the most striking features are the very low levels of
polarisation detected in each of the radio lobes with, in particular, no
polarised emission being detected from the northern half of the southern
lobe. Where significant polarisation is detected, the magnetic field
vectors are oriented parallel to the radio jet directions (see
Figure~2i). Figure~2e shows the depolarisation measure calculated between
8210 and 4710\,MHz; the depolarisation of the northern lobe is strong, but
relatively constant, whilst the southern lobe shows a steep gradient of
strengthening depolarisation towards the north. This is demonstrated in
Figure~\ref{368extra}a where the depolarisation measures between 8210 and
4710\,MHz, and between 4710 and 1420\,MHz at the same angular resolution,
are plotted as a function of declination throughout the southern lobe. It
is apparent that the depolarisation strengthens to the north. 

In the regions where polarised emission is detected, the rotation measure
has been calculated from the 4535, 4885 and 8210\,MHz observations as
described in Section~\ref{images}, and is shown in Figure~2f.  The global
rotation measures of the northern and southern lobes are 100 and
63\,rad\,m$^{-2}$ respectively. The northern radio lobe shows little
structure in the rotation measure map, but the southern lobe shows large
gradients in its rotation measure, mirroring the variations in
depolarisation measure: the rotation measure is relatively constant
throughout the southern half of the lobe (variations $\lta
20$\,rad\,m$^{-2}$), and then increases dramatically towards the
north--east and decreases dramatically towards the north--west. Because
the high and low rotation measures in the north--east and north--west
corners arise from regions where the polarisation is low, to confirm that
the values are real, in Figure~\ref{rotmes} we plot the polarisation
position angle against wavelength squared from 8\,GHz to 1.4\,GHz,
averaged throughout these two regions and through the `constant RM' area
in the south of the lobe. The position angles, allowing for $n\pi$
ambiguities, are well fit by a $\lambda$--squared law across all five
wavelengths in each of these three locations, with rotation measures of
88\,rad\,m$^{-2}$ for the southern region of the lobe (dotted line),
224\,rad\,m$^{-2}$ for the north--east corner (dashed line), and
$-$36\,rad\,m$^{-2}$ for the north--west corner (solid line). This
confirms the large variations in the rotation measure.

3C368 lies at moderately low galactic latitude ($b = 15^{\circ}$), and so
we cannot exclude a galactic origin for rotation measures of about
100\,rad\,m$^{-2}$. However, the large gradients in the rotation measure
and the correlation between the depolarisation and rotation measure
structures indicate that they are more likely to be associated with
Faraday rotation by matter local to 3C368. The measured rotation measures
must therefore be scaled up by a factor of $(1+z)^2$, or 4.55, giving mean
rest--frame rotation measures for the northern and southern lobes of 455
and 263\,rad\,m$^{-2}$ respectively, with the southern lobe rotation
measure changing from about $-160$ to about 1000\,rad\,m$^{-2}$ over a
distance of less than 10\,kpc. Furthermore, the large gradient in
depolarisation measure towards the northern end of the southern lobe
suggests that the lack of polarised emission at 8~GHz from this region of
the source is likely to be due this region of the source having been
strongly depolarised, and thus the intrinsic rotation measure values for
that region might be very large.

Using equation~\ref{deltaeqn}, we can estimate the expected dispersion in
rotation measure values if the depolarisation is associated with
variations in the Faraday depth of material within each beamwidth; this
gives $\Delta \sim 680$\,rad\,m$^{-2}$ for the southern lobe and $\Delta
\sim 360$\,rad\,m$^{-2}$ for the northern lobe, again comparable to the
observed values. The depolarisation expected between 4.7 and 1.4\,GHz by
Burn's law (equation~\ref{depollaw}) for these values of $\Delta$ is far
larger than observed. This is consistent with the slower fall--off of
depolarisation with wavelength predicted by Tribble \shortcite{tri91} for
an external Faraday screen with rotation measure variations on a scale
comparable to the beam size, and supports the hypothesis that the
depolarisation and Faraday rotation occur in a external Faraday screen
local to the radio source. 

\subsection{Radio--optical correlations}
\label{368radopt}

In Figures~2g and h, the radio emission of 3C368 is compared
with the optical and infrared emission. Figure~2g shows total
intensity 5\,GHz radio contours overlaid upon the grey--scale HST image of
the source presented by Longair \etal\ \shortcite{lon95}. This image is a
combination of two 30-minute HST exposures taken through the F702W and
F791W filters. Figure~2h shows the same radio contours overlaid
upon an infrared `H+K' image kindly provided by Alan Stockton; this is a
7200 second exposure through a combined H and K--band filter taken with
the University of Hawaii 2.2m telescope, with an angular resolution of
0.32 arcseconds.

\begin{figure*}
\begin{tabular}{cc}
\psfig{figure=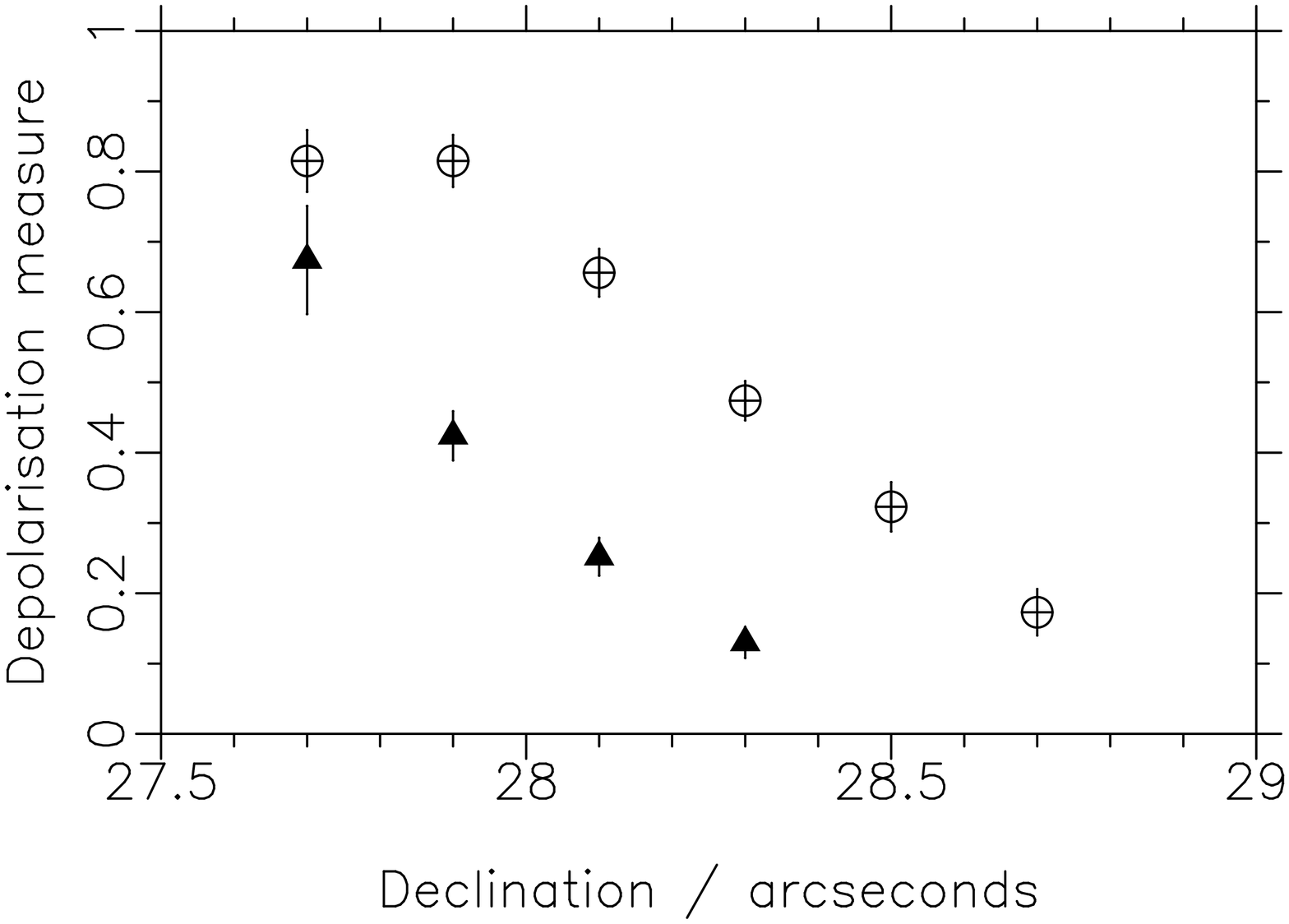,clip=,angle=-90,width=8cm}
&
\psfig{figure=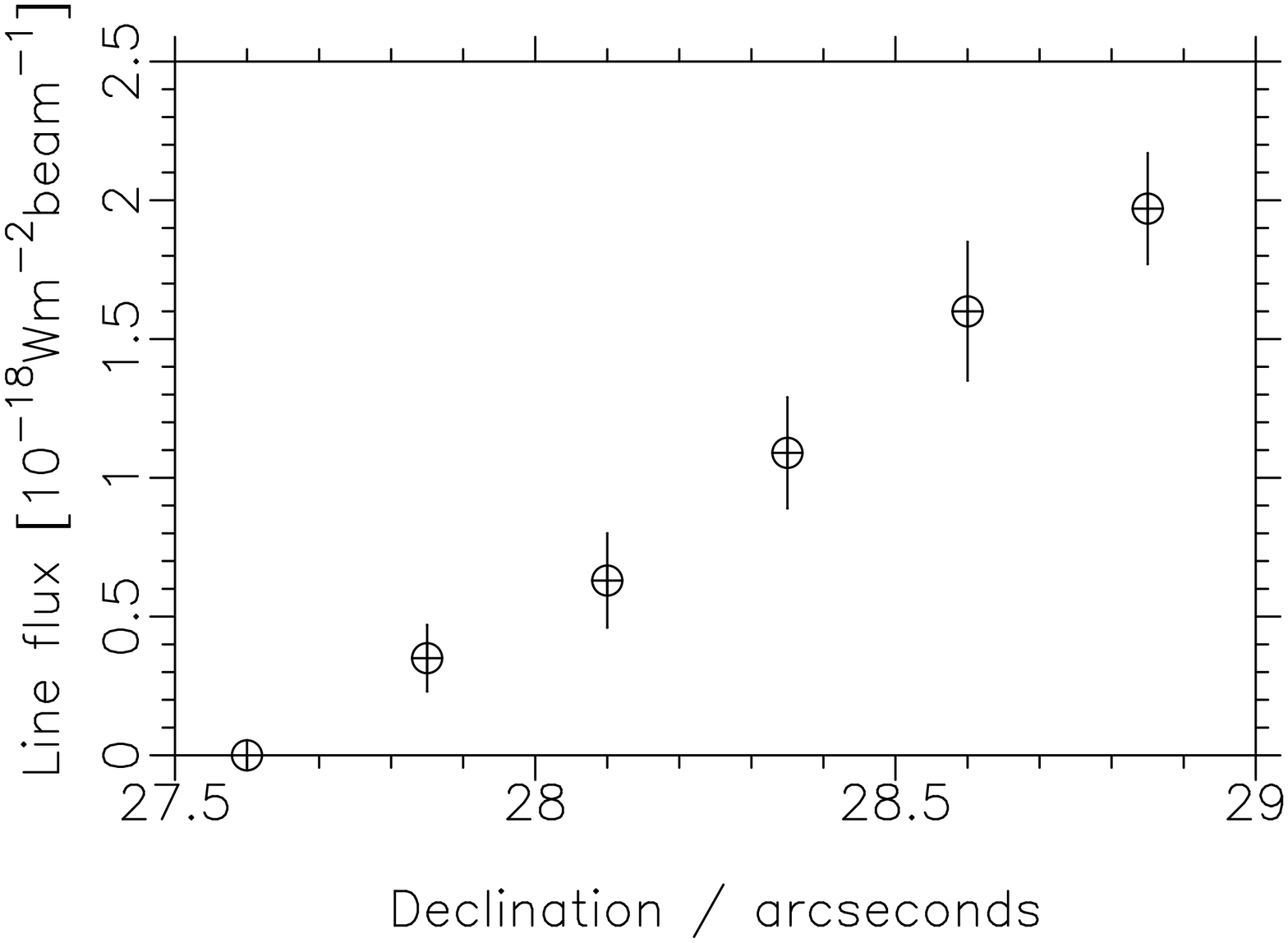,clip=,angle=-90,width=8cm}
\end{tabular}
\caption{\label{368extra} (a) The variation of depolarisation measure with
position through the southern radio lobe of 3C368, going from south to
north. The open circles show the 8210 to 4710\,MHz depolarisation measure,
and the filled triangles show the 4710 to 1420\,MHz values. (b) The
variation of the [OII] emission line flux (as measured in the observations
of Meisenheimer and Hippelein 1992) over the same declination range.}
\end{figure*}

The nature of the HST and infrared images has been extensively discussed
elsewhere \cite{lon95,sto96a,bes97c}, and we do not propose to repeat that
here. We note only that the bright unresolved emission towards the centre
of the HST image (RA 18 05 06.39, Dec 11 01 31.3) is a foreground galactic
M--dwarf star, and is thus unrelated to the structure of the radio galaxy
\cite{ham91}. Assuming that the northern radio knot is the radio core, we
align this with the central peak of the infrared emission from the galaxy,
clearly visible on Figure~2h, about 1.5$''$ north of the M--star. The
astrometry should be accurate to about 0.1 arcsec. The infrared and HST
images were astrometrically aligned using the position of the M--star.  A
number of features are now apparent from the optical and infrared images:

\begin{enumerate}
\item As noted by Stockton \etal\ \shortcite{sto96a}, the centre of the
infrared image of the galaxy coincides with a minimum in the HST image. We
have investigated the possibility that this may be associated with a dust
lane obscuring the central regions, as is frequently observed in 3CR radio
galaxies at lower redshifts \cite{dek96}. A model galaxy was constructed
with a K--magnitude equivalent to that of 3C368, K$= 17.03 \pm 0.15$, and
a colour of R$-$K $= 5.3 \pm 0.2$, corresponding to the range expected for
a passively evolving galaxy at this redshift which formed at some redshift
$z_{\rm f} > 3$ \cite{bru93}. This galaxy was constructed to have a radial
intensity profile following de Vaucouleurs' law, with a characteristic
radius given by the mean of those determined for the distant 3CR galaxies,
corresponding to $r_{\rm e} = 1.7 \pm 0.4$ arcsec
\cite{bes98d}\footnote{Note that due to the presence of the M--star, the
characteristic radius of 3C368 itself could not be determined in the work
of Best \etal}. The flux density expected for this model galaxy through a
central 0.3$''$ diameter aperture was then compared with that measured
from the HST observations. We find that extinction with $E(B-V) = 0.27 \pm
0.15$ is required to obscure the central regions of the galaxy. Although
only marginally significant, this value is comparable to that derived by
Dickinson \etal\ \shortcite{dic96} for 3C324. This value be higher if some
of the flux density observed from this region is associated with the
aligned emission rather than the old stellar population, as is suggested
by the fairly uniform level of emission from the extended region
underlying the bright optical knots.

\item The radio jet extending towards the northern lobe is coincident with
the northern string of bright optical knots of the HST image. Indeed,
there is a continuous optical feature running from the radio nucleus
towards the hotspot. This feature undergoes a slight bend near RA 18 05
06.40, Dec 11 01 34.2, close to a feature clearly visible in the infrared
image of the galaxy. This object is significantly redder than the optical
jet, and is likely to be a companion galaxy to the host of 3C368
\cite{rig92}. An interaction of the radio jet with this object may have
been responsible for the brightening of the jet, as well as for the intense
nebular continuum emission from this region
\cite{dic95,sto96a}.

\item Polarised radio emission is only detected in the region of the
southern radio lobe which does not overlap with the bright optical
emission; the southern hotspot shows little depolarisation, even between
1.4 and 5\,GHz (see Figure~\ref{368extra}a), whilst the region of the
southern lobe close to the boundary of the optical emission shows very
strong depolarisation and high rotation measure gradients. No polarised
emission is detected from the northern half of this lobe, even at
8\,GHz. A radial decrease in the strength of the depolarisation is
expected if the radio source is sitting in an X--ray halo \cite{gar91},
although for the rapid decline in depolarisation seen here, the density
profile of the halo would have to have a very small scale size, of order
30\,kpc. Alternatively, the depolarisation could involve gas associated
with the radio galaxy itself; a significant proportion of the southern
optical emission region of 3C368 is associated with emission line gas
\cite{mei92,rig92,sto96a}, and large gradients in the Faraday rotation by
this gas may have been responsible for depolarising of the northern half
of the southern radio lobe. In Figure~\ref{368extra}b, we display the
variation of the line flux of the [OII]~3727 emission line with
declination, through the region of the southern lobe in which the
depolarisation measure changes rapidly. These data are taken from
Meisenheimer and Hippelein \shortcite{mei92}, corrected for the improved
astrometry of the optical reference frames. The location of the emission
line gas correlates directly with the region of strong depolarisation.

If the emission line gas is responsible for this depolarisation, then it
must have a covering fraction, $f_{\rm c}$, of close to unity, which
allows some characteristics of the emission line clouds to be
determined. Assuming that the emission line clouds have temperatures $T
\sim 10^4$\,K and are in pressure equilibrium with the hot gas in which
they are embedded, they will have densities of a few times $10^7$ to a few
times $10^8$\,m$^{-3}$. McCarthy \shortcite{mcc88} showed that, in the low
density limit, the luminosity of the [OII]\,3727 emission line could be
related to the gas density and volume by the equation $L_{\rm 37} = 5.3
\times 10^{-11} n_{\rm e}^2 f_{\rm v} V_{\rm 64}$, where $L_{\rm 37}$ is
the luminosity of the [OII]\,3727 line in units of $10^{37}$\,W, $n_{\rm
e}$ is the electron density in m$^{-3}$, $f_{\rm v}$ is the volume filling
factor and $V_{\rm 64}$ is the total volume in units of
$10^{64}$\,m$^{3}$. For 3C368, $L_{\rm 37} \approx 0.8$ (Meisenheimer and
Hippelein 1992, corrected for our adopted cosmology)\nocite{mei92}. The
HST image indicates that the emission occurs from a region on the sky
subtending about 7 by 2 arcseconds; assuming cylindrical symmetry, $V_{\rm
64} \sim 0.04$. The above equation therefore gives a volume filling factor
of between about $3 \times 10^{-4}$ and $3 \times 10^{-6}$, which
corresponds to a total mass of warm gas of a few times $10^8$ to a few
times $10^9 M_{\odot}$, comparable to estimates in other radio galaxies
\cite{mcc93,rot96d}. An upper limit to the sizes of the individual
emission line clouds, $d$, can then be calculated by $d \sim l f_{\rm v} /
f_{\rm c}$, where $l \sim 17$\,kpc is the size of the line emitting
region. This upper limit is between 0.05 and 5\,pc (cf. the cloud sizes of
0.035\,pc derived in extended \la\ halos by van Ojik \etal\
1997).\nocite{oji97}

The rotation measure is related to the gas density by the equation $RM =
8.1 \times 10^{-3} \int_0^l n_{\rm e} B_{\parallel} {\rm d}l$, where $RM$
is measured in rad\,m$^{-2}$, $n_{\rm e}$ in m$^{-3}$, $B_{\parallel}$ in
nT and $l$ in kpc \cite{bur66}. From the above calculation, the path
length through the warm emission line clouds would be a factor of between
$10^4$ and $10^6$ lower than the path length through the surrounding hot
phase gas, whilst the densities of these clouds are only a factor of
$10^3$ to $10^4$ higher. Especially considering that the net Faraday
rotation from an ensemble of cloudlets within each synthesised radio beam
will get averaged down, for the warm emission line gas to dominate the
rotation measure, the magnetic fields within the cloudlets would have to
be factors of at least 10 to 1000 higher than those in the surrounding hot
gas phase (usually estimated at around 1\,nT, e.g. Carilli \etal\
1997).\nocite{car97}

Thus, if the emission line gas is responsible for depolarising the
southern radio lobe of 3C368 then the emission line clouds must have a
covering fraction of unity which restricts their sizes to parsec or
subparsec scales, {\it and} the condensation of these emission line
cloudlets out of the hot gas must be accompanied by a strong amplification
in the magnetic field strengths. The alternative is that the
depolarisation and Faraday rotation are associated with the hot cluster
gas. In this case either the scale size of the X--ray halo gas would have
to be co-incidently the same size as the extent of the emission line gas,
or the emission line gas must act directly as a tracer for the region of
enhanced density in the hot gaseous phase.

\begin{figure}
\psfig{figure=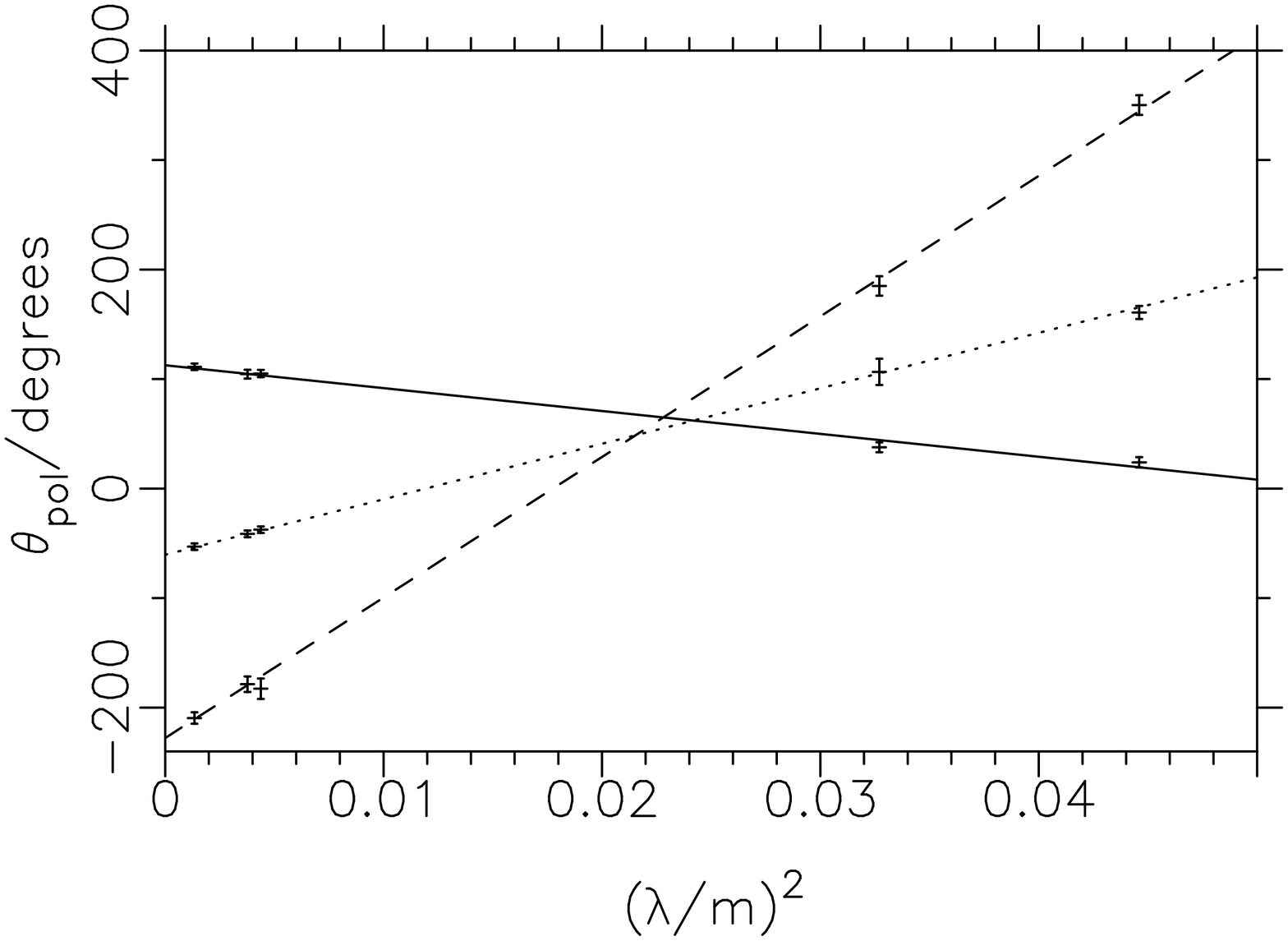,clip=,angle=-90,width=7.8cm}
\caption{\label{rotmes} Fits to the polarisation position angle, allowing
for $n\pi$ ambiguities, versus $\lambda^2$ relation in three regions of
the southern radio lobe of 3C368: the southern region (dotted line), the
north--east corner (dashed line) and the north--west corner (solid line).}
\end{figure} 

\item The southern region of emission on the HST image shows an elliptical
structure with a central minimum. Meisenheimer and Hippelein
\shortcite{mei92} argued that the velocity structures of the emission line
gas appear consistent with this region being associated with a central bow
shock caused by the radio hotspot.  With our improved astrometry, for
which the radio--optical correlations in the previous points have provided
further support, it is apparent that the southern radio hot--spot cannot
coincide with this `hole' in HST emission.  The emission line velocities
are not associated with a bow--shock at the current position of the
hotspot of the radio emission, and there is also no convincing evidence in
the radio source structure for any precession of the radio source axis,
whereby the radio hotspot may have advanced through this region at an
earlier time. If this is truly a bow--shock structure then its origin is
unclear.

\item It is important to note that the unresolved southern radio knot is
{\it not} coincident with the position of the M--star. Many dwarf M--stars
display variable radio emission \cite{whi89,spe93}, and Hammer \etal\
\shortcite{ham91} noted that the 5\,GHz flux density of this radio knot
differs between the 0.45\,mJy, measured by Djorgovski \etal\
\shortcite{djo87}, and the 3.2 to 6.4\,mJy on the radio map of Chambers
\etal\ \shortcite{cha88}. They argued that this radio emission might be
associated with the star. This flux density `variation' is, in reality,
due to an error in the contour labelling on the Chambers \etal\ map: their
lobe fluxes are also a factor of ten greater than those measured on our
map. The 5\,GHz flux density that we measure for the southern knot is
0.55\,mJy, comparable with the value of Djorgovski et~al. Furthermore, for
a typical dwarf M--star at a distance of 500\,pc, that derived by Hammer
\etal\ based upon the luminosity and colour of the star, the radio flux at
5\,GHz even during flares would generally not exceed a few $\mu$Jy, and
the quiescent value would be much lower. We conclude that this southern
knot is part of the radio source 3C368 rather than arising from the
M--star.
\end{enumerate}

\subsection{The emission line gas}
\label{368gas}

The velocity structure of the emission line gas of 3C368
\cite{mei92,sto96a} is indicative of either infall or outflow along the
radio axis, depending upon which lobe of the source is closer to
us. Naively, the radio properties of the source can be used to infer the
orientation of the axis relative to the line of sight. Stronger
depolarisation is observed in the southern lobe as compared with the
northern lobe, which may be interpreted in terms of the Laing--Garrington
effect \cite{lai88,gar88} to imply that the southern lobe is the more
distant, seen through the greater depth of material. Such a conclusion
would also be consistent with the detection of a `Doppler--boosted'
one--sided radio jet in the northern arm of the source. This result would
imply that the emission line gas is infalling towards the central galaxy.

On the other hand, although the Laing--Garrington effect is almost
universal in quasars, both Pedelty \etal\ \shortcite{ped89a} and McCarthy
\etal\ \shortcite{mcc91} find that in their samples of radio galaxies the
depolarisation depends more upon local environmental effects than upon
orientation. We have already argued that the emission line gas associated
with the optically bright regions overlying the southern lobe may be
responsible for much of the depolarisation in this region. In addition,
the northern lobe is strongly depolarised between 5\,GHz and 1.4\,GHz (see
Table~\ref{radprops}). Furthermore, this source is not a classic
jet\,/\,counter--jet example: the jet in the northern arm is weak whilst
the second central knot lies in the southern arm and the southern lobe is
elongated and knotty. The emission from the radio jet in the northern arm
is more likely to have been brightened through its interactions with the
interstellar medium than by beaming effects. We conclude, therefore, that
the radio properties are most likely dominated by environmental effects
and cannot be used to indicate the orientation of 3C368.

The tight correlation seen between the locations of bright optical and
radio emission, which implies that interactions between the radio jet and
its environment are of great importance in this source, argues against an
infall scenario, since any infalling material would be greatly
decelerated by the kinetic thrust of the radio jet. More likely is that
the jet--cloud interactions would accelerate the emission line gas
outwards, and produce the observed velocities as outflows \cite{sto96a}.

\section{Discussion}
\label{concs} 

We summarise below the main observational results of this work:

\begin{enumerate} 
\item We have detected radio core candidates in each of 3C324 and
3C368. Aligning these relatively flat spectrum knots with the centre of
the infrared images enables the radio and the optical\,/\,infrared frames
of references to be astrometrically aligned to an accuracy of about 0.1
arcsec.

\item The radio cores of each source lie at a minimum of the optical
emission, probably associated with obscuration by a central dust
lane. Dust may also play an important role in determining the extended
morphology of the optical emission.

\item Radio jets are detected in both sources, and in each case are
co-linear with the strings of bright optical knots.

\item These observations sort out the disparity between lower angular
resolution observations of the polarisation structures of these radio
galaxies \cite{ped89a,fer93}. The higher angular resolution reveals very
large gradients in the rotation measure structures, reaching up to
1000\,rad\,m$^{-2}$ over a distance of order 10\,kpc, and regions of
strong depolarisation.

\item There is a tight correlation between the depolarisation measures and
the gradients in the rotation measure, suggesting that the depolarisation
is external, and caused by variations within each beamwidth of the Faraday
depth of material in the vicinity of the radio source.
\end{enumerate}

The striking co-linearity of the radio jet with the bright optical knots
in the two sources indicate that interactions of the radio jet itself,
rather than just the AGN or the radio cocoon, must play a critical role in
the origin of the aligned optical emission. Of interest is that, given
these strong interactions between the radio jets and their environment,
more enhanced radio emission is not seen in these regions, in contrast to
that observed in bright radio knots at the sites of jet--cloud
interactions at low redshifts (e.g. 4C29.30, van Breugel \etal\
1986). This suggests that the more powerful radio jets in these high
redshift sources are not as strongly disrupted by their
interactions. 

Despite the many similarities between the two sources, there are equally
important differences. The northern arm of 3C368 shows a tight correlation
between the regions of radio and optical emission, whilst in 3C324
although these are co-linear they are not co-spatial. The nature of the
optical (rest--frame ultraviolet) emission of the two galaxies is also
very different. 3C324 has a high percentage of spatially extended optical
polarisation, $P \approx 11\%$, indicating that illumination by the
central AGN must be important in this source \cite{cim96}. In contrast,
HST and Keck observations of 3C368 have detected no polarised optical
emission \cite{bre96c}.

3C368 is an exceptional source, being probably the most aligned of all of
the 3CR radio galaxies and possessing the highest emission line flux. The
extent of its radio emission is 73\,kpc, only slightly greater than that
of the optical emission. Its structure may be interpreted in terms of it
being a relatively young radio source (a few times $10^6$ years), in which
on--going interactions between the radio jet and the interstellar medium
remain the dominant effects. In the northern region of this source, where
the bright optical knots lie coincident with the radio jet, the nebular
continuum emission is very luminous and the emission line velocities are
high \cite{dic95,sto96a}.  Interactions between the radio jet and its
environment, possibly including a companion galaxy (see
Section~\ref{368radopt}) may brighten the radio jet and accelerate and
shock--ionise the warm emission line gas, explaining all of these
features. The strong correlation between the radio and optical structures
cannot be explained if the gas responsible for the nebular continuum is
photoionised by the AGN. Jet--induced star formation in this region cannot
be excluded, but there is no direct evidence for this. Any small
contribution of scattered light might not be detectable due to the
dilution of the polarisation by the very strong nebular continuum
contribution.

The radio emission of 3C324 is more extended and reaches well beyond the
region of bright optical emission. In this source, we may be observing the
residual effects after the radio jets have forced a passage through the
host galaxy. In the aftermath of the jet shocks the nebular continuum
contribution has decreased, and now some scattered light is
detected. Cimatti \etal\ \shortcite{cim96} calculate that at most 30 to
50\% of the optical flux density can be associated with scattered
radiation, with the rest likely to be nebular continuum emission or light
from a young starburst. They suggest that the knot close to the western
lobe may currently be undergoing a burst of star formation at a rate of
$70 M_{\odot}$\,yr$^{-1}$. Most of the optical emission from 3C324 arises
from the bright knotty structures along the radio axis, with no evidence
for the scattered emission being evenly distributed over an ionisation
cone structure. Therefore, the scattering material in this source must
also be preferentially located along the jet direction.  Two possible
mechanisms for this latter effect would the production of dust in a region
of star formation induced by the radio jet, or the disintegration of
cooled clumps of optically thick gas by the jet, and the exposure of
previously hidden dust grains at their centre \cite{bre96b}.

\smallskip

At low redshifts, rotation measure values and variations in excess of a
few hundred rad\,m$^{-2}$ are seen in two types of radio source, compact
steep spectrum radio sources in which the radio lobes lie within the host
galaxy (e.g. Taylor \etal\ 1992)\nocite{tay92}, and FR\,II radio sources
located in clusters with luminous X--ray cooling flows. For the former,
Garrington and Akujor \shortcite{gar96} have suggested that the rotation
measure correlates with linear size, with the dense gas located towards
the centre of the host galaxy being responsible for the Faraday rotation.
For the latter, Taylor \etal\ \shortcite{tay94} found a correlation
between the rotation measure and the cooling flow rate, suggesting that
the intracluster medium is responsible.

We have argued that the large rotation measures observed in the southern
lobe of 3C368 are due to emission line gas associated with the host
galaxy. The northern lobe of this source, however, and both lobes of
3C324, lie distant from the host galaxy. Dickinson \shortcite{dic97a} has
reported the detection of extended X--ray emission at a luminosity of
$L_{\rm X} = (8.1 \pm 1.6) \times 10^{44}$\,erg\,s$^{-1}$ from 3C324, a
value comparable to that of the Coma cluster. He associates this with
cooling intracluster gas. Similarly, 3C368 has been detected using the
ROSAT PSPC, with a signal--to--noise ratio of 5, giving $L_{\rm X} \sim
1.7 \times 10^{44}$\,erg\,s$^{-1}$ \cite{cra95}. If these X--ray
luminosities are associated with cooling flows, then the cooling flow
rates and rotation measures lie roughly upon the correlation of Taylor
\etal\ \shortcite{tay94}, suggesting that the high rotation measures of
3C324 and of the regions of 3C368 away from the emission line gas are
indeed associated with their intracluster gas. These rotation measure
values are not as extreme as, for example, the 20000\,rad\,m$^{-2}$ seen
in 3C295, at the centre of a very rich cluster of galaxies at $z = 0.461$
\cite{per91}, but the large values and gradients provide further support
for the hypothesis that powerful distant radio galaxies lie in at least
moderately rich young cluster environments.

\section*{Acknowledgements} 

The National Radio Astronomy Observatory is operated by Associated
Universities Inc., under co-operative agreement with the National Science
Foundation. MERLIN is operated by the University of Manchester on behalf
of PPARC. The NASA/ESA Hubble Space Telescope observations were obtained
at the Space Telescope Science Institute, which is operated by Associated
Universities Inc., under contract from NASA. The United Kingdom InfraRed
Telescope is operated by the Joint Astronomy Centre on behalf of PPARC.
We thank Alan Stockton for kindly providing a digital version of the
infrared data for 3C368. This work was supported in part by the Formation
and Evolution of Galaxies network set up by the European Commission under
contract ERB FMRX--CT96--086 of its TMR programme. We thank the referee,
Clive Tadhunter, for some helpful comments.

\label{lastpage}
\bibliography{pnb} 
\bibliographystyle{mn} 

\end{document}